\documentstyle[12pt]{article}
\makeatletter
\def\appendix{\par\clearpage
  \setcounter{section}{0}
  \setcounter{subsection}{0}
  \@addtoreset{equation}{section}
  \def\@sectname{Appendix~}
  \def\theequation{\thesection.\arabic{equation}}
  \def\thesection{\Alph{section}}}
\makeatother
\begin{document}
\begin{titlepage}
\hskip 11cm \vbox{
\hbox{Budker INP 2000-62}
\hbox{UNICAL-TH 00/2}
\hbox{July 2000}}
\vskip 0.3cm
\centerline{\bf The Compatibility of the Gluon Reggeization with the $s$-channel Unitarity$^{~\ast}$}
\vskip 0.8cm
\centerline{  V.S. Fadin$^{a, b~\dagger}$, R. Fiore$^{c, d~\ddagger}$, M.I. Kotsky$^{d~\ddagger}$\footnote
{On leave of absence from the Budker Institute for Nuclear Physics, Novosibirsk, Russia.}}
\vskip .3cm
\centerline{\sl $^a$ Budker Institute for Nuclear Physics, 630090 Novosibirsk, Russia}
\centerline{\sl $^b$ Novosibirsk State University, 630090 Novosibirsk, Russia}
\centerline{\sl $^c$ Dipartimento di Fisica, Universit\`a della Calabria,}
\centerline{\sl  I-87036 Arcavacata di Rende, Cosenza, Italy}
\centerline{\sl $^d$ Istituto Nazionale di Fisica Nucleare, Gruppo collegato di Cosenza,}
\centerline{\sl I-87036 Arcavacata di Rende, Cosenza, Italy}
\vskip 1cm
\begin{abstract}
Recently the non-forward BFKL kernel for interaction of two Reggeized
gluons in the antisymmetric colour octet state in the $t$-channel was
obtained in the next-to-leading order. It gives the possibility to check in
this order the bootstrap condition for this kernel, appearing as the
requirement of the compatibility of gluon Reggeization with the
$s$-channel unitarity.
\end{abstract}
\vfill
\hrule
\vskip.3cm
\noindent
$^{\ast}${\it Work supported in part by the Ministero italiano dell'Universit\`a e della Ricerca Scientifica e Tecnologica,
in part by INTAS and in part by the Russian Fund of Basic Researches.}
\vfill
$ \begin{array}{ll}
^{\dagger}\mbox{{\it e-mail address:}} &
 \mbox{FADIN ~@INP.NSK.SU}\\
\end{array}
$
\vfill
\vskip -1.5cm
$ \begin{array}{ll}
^{\ddagger}\mbox{{\it e-mail address:}} &
 \mbox{FIORE, KOTSKY ~@FIS.UNICAL.IT}\\
\end{array}
$
\vfill
\vskip .1cm
\vfill
\end{titlepage}
\eject

\section{ Introduction}

Reggeization of elementary particles in the non-Abelian gauge
theories~\cite{GriSch},~\cite{Lip},~\cite{FaSher},
particularly in QCD, seems to be one of the most intriguing properties of
these theories. Origin and consequences of this property are not completely
understood, but its role in the high energy behaviour of scattering
amplitudes hardly can be overestimated. The gluon Reggeization determines
the high energy behaviour of non-decreasing with energy cross sections in
perturbative QCD. In particular, it appears to be the basis of the BFKL
approach~\cite{BFKL}.

Originally the BFKL equation was derived in the leading logarithm
approximation (LLA), that means summation of all terms of perturbation
theory, where smallness of the coupling constant $\alpha_s$ is
compensated by the large logarithm $\ln s$ of the squared c.m.s. energy. Two
years ago the first radiative correction to the kernel of the BFKL equation
for the forward scattering, i.e. $t=0$ and colour singlet state in the $t$-channel,
was found~\cite{FL98}. It gives the possibility to sum
also the next-to-leading order (NLO) terms. The correction was derived
assuming the validity of the gluon Reggeization in the NLO, which is not
proved, and therefore must be carefully tested. That can be done by checking
the bootstrap equations~\cite{FF} appearing as the conditions of
the compatibility of the gluon Reggeization with the $s$-channel unitarity.
The second, not less important reason for this check is that it gives the
over-crossed test of the calculations of almost all ingredients used in the BFKL
approach. In this approach the high energy scattering amplitudes are given
by the convolution of the impact factors of scattered particles and the
Green function for the Reggeon-Reggeon scattering. The impact factors and
the kernel of the BFKL equation for the Green function, in turn, are
expressed through the effective vertices for the Reggeon-particle
interactions and the gluon Regge trajectory. The first bootstrap equation
ties the kernel of the BFKL equation for the antisymmetric colour octet
state of the two Reggeized gluons in the $t$-channel ${\cal K}^{(8)}\left( \vec q_1, \vec q_2 ; \vec q \right)$
with the gluon
trajectory $j=1+\omega (t)$. In the NLO the equation takes the form 
\begin{equation}
\frac{g^{2}Nt}{2\left( 2\pi \right) ^{D-1}}\int \frac{d^{D-2}q_{1}}{\vec{q}%
_{1}^{\,2}\vec{q}\,_{1}^{\prime \,2}}\int \frac{d^{D-2}q_{2}}{\vec{q}%
_{2}^{~2}\vec{q}_{2}^{\,\prime ~2\,}}{\cal K}^{(8)(1)}\left( \vec q_1, \vec q_2 ; \vec q \right)
=\omega ^{\left(
1\right) }\left( t\right) \omega ^{\left( 2\right) }\left( t\right)
\label{z1}
\end{equation}
where $g$ is the coupling constant, $N$ is the number of colours ($N=3$ for
QCD), $D=4+2\epsilon $ is the space-time dimension taken different from 4
for regularization of infrared divergences. ${\cal K}^{(8)(1)}\left( \vec q_1, \vec q_2 ; \vec q \right)$
 is the first radiative
correction to the octet kernel, the vector sign is used for denoting
component momenta transverse to the plane of momenta of initial particles, $%
\vec{q}^{~}$is the momentum transfer in the scattering process. For the sake
of brevity we put $\vec{q}\,_{i}^{\prime \,}=\vec{q}_{i}^{~}-\vec{q}$, $%
i=1,2 $; \ $\vec{q}_{1}^{~}$ and $-\vec{q}\,_{1}^{\prime \,}$\ (\ $\vec{q}%
_{2}^{~}$ and $-\vec{q}\,_{2}^{\prime \,}$) being the transverse momenta of
the initial (final) Reggeized gluons in the $t$-channel.

The second bootstrap equation connects the colour octet impact factor of any  particle
and its effective interaction vertex with the Reggeized
gluon. This equation was checked for the gluon~\cite{FFKP1} and
quark~\cite{FFKP2} impact factors. It was proved that the equation
is fulfilled both for helicity conserving and non-conserving parts of the impact factors at
arbitrary space-time dimension $D$. As for the first bootstrap equation,
Eq. (\ref{z1}), it was checked (and was found to be satisfied also
at arbitrary $D$) only in the quark part of this equation~\cite{FFP}.
The most important (and much more complicated) gluon part of this
equation remained unchecked till now. Recently, the gluon part of the
radiative correction to ${\cal K}^{\left( 8\right)}$ was obtained~\cite{FG},
that gives a possibility to analyze Eq. (\ref{z1}) in this part.

In this paper we check the first bootstrap condition (\ref{z1}).
Since the fulfillment of this equation for the quark part is
established~\cite{FFP}, we consider pure gluodynamics. In the next
section we present the explicit expressions for the octet BFKL kernel
and the gluon trajectory, and analyse Eq. (\ref{z1}). The proof of
its fulfillment is given in Section 3. In the last section we discuss possible generalizations.

\section{Explicit Form of the Bootstrap Equation}

Let us start with the gluon trajectory. Its one-loop expression~\cite{Lip} is well known:
\begin{equation}
\omega^{(1)}(t)=\frac{g^{2}Nt}{2\left( 2\pi \right) ^{D-1}}\int \frac{%
d^{D-2}k}{\vec k^2(\vec q - \vec k)^2}~.
\label{z2}
\end{equation}
The integral can be easily calculated at arbitrary $D$ and one has 
\begin{equation}
\omega ^{(1)}(t)=-\bar{g}^{2}(\vec q^{~2})^{\epsilon }\frac{\Gamma ^{2}(\epsilon )%
}{\Gamma (2\epsilon )}~,
\label{z3}
\end{equation}
where $\Gamma $ is the Euler gamma-function and
\begin{equation}
\bar{g}^{2}=\frac{g^{2}N\Gamma (1-\epsilon )}{(4\pi )^{D/2}}~.
\label{z4}
\end{equation}
We stress that everywhere in this paper we use the unrenormalized
coupling constant $g$.

In the two-loop approximation the integral representation for the
trajectory is~\cite{F95} 
\begin{equation}
\omega^{(2)}(t)=\frac{g^{2}t}{(2\pi )^{D-1}}\int \frac{d^{D-2}q_{1}}{%
\vec{q}_{1}^{\,2}\vec{q}_{1}^{\,^{\prime }2}}\left[ F_{G}(\vec{q}_{1}^{\,},%
\vec{q}\,)-2F_{G}(\vec{q}_{1}, \vec{q}_{1})\right] ~, 
\label{z5}
\end{equation}
where
$$
F_{G}(\vec{q}_{1}^{\,},\vec{q}\,)=-\frac{g^{2}N^{2}\vec{q}^{\,2}}{4\left(
2\pi \right) ^{D-1}}\int \frac{d^{D-2}q_{2}}{\vec{q}_{2}^{~2}\left( \vec{q}%
_{2}^{\,\,}-\vec{q}\right) ^{2}}\left[ \ln\left( \frac{\vec{q}^{\,2}}{%
\left( \vec{q}_{1}^{\,}-\vec{q}_{2}^{\,}\right) ^{2}}\right) -2\psi \left(
1+2\epsilon \right) \right.
$$
\begin{equation}
\left. -\psi \left( 1-\epsilon \right) +2\psi \left( \epsilon \right) +\psi
\left( 1\right) +\frac{1}{1+2\epsilon }\left( \frac{1}{\epsilon }+\frac{%
1+\epsilon }{2\left( 3+2\epsilon \right) }\right) \right]~,
\label{z6}
\end{equation}
and $\psi (x)=\Gamma ^{\prime }(x)/\Gamma (x)$.
The integral (\ref{z5}) can be expressed in terms of elementary functions
only for $\epsilon \rightarrow 0$. The answer is~\cite{FFK}
\begin{equation}
\omega^{(2)}(t) \simeq \left( \frac{{\bar g}^{2}\left( \vec{q}^{\,2}\right) ^{\epsilon }%
}{\epsilon }\right) ^{2}\left[ \frac{11}{3}+\left( 2\psi ^{\prime }(1)-\frac{%
67}{9}\right) \epsilon +\left( \frac{404}{27}+\psi ^{\prime \prime }(1)-%
\frac{22}{3}\psi ^{\prime }(1)\right) \epsilon ^{2}\right]~.
\label{z7}
\end{equation}

The kernel of the BFKL equation for any of the colour states ${\cal R}$
of two Reggeized gluons in the $t$-channel is given by the sum of the
``virtual'' part, defined by the gluon trajectory, and\ the ``real'' part
${\cal K}_{r}^{\left({\cal R}\right)}$, related to the real particle
production:
\begin{equation}
{\cal K}^{\left({\cal R}\right)}\left( \vec{q}_{1},\vec{q}_{2};\vec{q%
}\right) =\left( \omega \left( -\vec{q}_{1}^{\,2}\right) +\omega \left( -%
\vec{q}\,_{1}^{\prime \,2}\right) \right) \vec{q}_{1}^{\,2}\vec{q}%
\,_{1}^{\prime \,2}\delta^{(D-2)} \left( \vec{q}_{1}^{\,}-\vec{q}_{2}^{\,}\right) +%
{\cal K}_{r}^{\left({\cal R}\right)}\left( \vec{q}_{1},\vec{q}_{2};%
\vec{q}\right)~.
\label{z8}
\end{equation}
The octet kernel in the Born approximation differs only by the coefficient
1/2 from the singlet (Pomeron) kernel: 
\begin{equation}
{\cal K}_{r}^{\left(8\right)\left(B\right) }\left( \vec{q}_{1},%
\vec{q}_{2};\vec{q}\right) =\frac{g^{2}N}{2\left( 2\pi \right) ^{D-1}}f_B~,\ \ \
f_B=\frac{\vec q_1^{~2}\vec q_2^{~\prime 2} + \vec q_2^{~2}\vec q_1^{~\prime 2}}{\vec k^{~2}}
- \vec q^{~2}~,
\label{z9}
\end{equation}
where $\vec{k}=\vec{q}_{1}^{\,}-$ $\vec{q}_{2}^{\,}$. The radiative
correction to this kernel is known only in the limit $\epsilon \rightarrow 0$~\cite{FG}.
It can be presented as the sum of three contributions: 
\begin{equation}
{\cal K}_{r}^{\left( 8\right) \left( 1\right) }\left( \vec{q}_{1},\vec{q}%
_{2};\vec{q}\right) =\frac{\bar{g}^{4}}{\pi ^{1+\epsilon }\Gamma (1-\epsilon
)}\left( {\cal K}_{1}+{\cal K}_{2}+{\cal K}_{3}\right)~.
\label{z10}
\end{equation}
The first contribution is proportional to the Born kernel (\ref{z9}): 
\begin{equation}
{\cal K}_{1}=-f_{B}\frac{(\vec{k}^{\,2})^{\epsilon }}{\epsilon }\left[ 
\frac{11}{3}+\left( 2\psi ^{\prime }(1)-\frac{67}{9}\right) \epsilon +\left( 
\frac{404}{27}+7\psi ^{\prime \prime }(1)-\frac{11}{3}\psi ^{\prime
}(1)\right) \epsilon ^{2}\right]~.
\label{z11}
\end{equation}
Note that this part of the kernel contains all explicit singularities in $%
\epsilon $, as it should be, because of factorization of singularities in
QCD amplitudes. Note also that in Eq. (\ref{z11}) the term $(\vec k^2)^\epsilon$
is not expanded in $\epsilon $ and the terms of order $\epsilon $ are
kept in the coefficient. It is done because the kernel is singular at $\vec{%
k}=0$, so that, at subsequent integrations of the kernel in the BFKL
equation, the region of arbitrary small, for $\epsilon \rightarrow 0$, values
of $\vec{k}^{\,2}$, where $\epsilon|\ln\vec{k}^{\,2}| \sim 1$,
does contribute and, moreover, leads to the appearance of an extra 1/$%
\epsilon $ factor. Consequently, the terms $\sim \epsilon $ are kept in the
coefficient to save all non-vanishing for $\epsilon \rightarrow 0$ terms
after the integration. The last two terms in Eq. (\ref{z10}) are known for $\epsilon \rightarrow 0$ only.
To stress this circumstance we shall indicate them with the superscript $^{(0)}$.
The second contribution can be expressed in terms of logarithms: 
$$
{\cal K}_{2}^{(0)}=\left\{ \vec q^{~2}\left[ \frac{11}{6}\ln
\left( \frac{\vec q_1^{~2}\vec q_2^{~2}}{\vec q^{~2}\vec k^{~2}}
\right) + \frac{1}{4}\ln\left( \frac{\vec q_1^{~2}}{\vec q^{~2}}
\right)\ln\left( \frac{\vec q_1^{~\prime 2}}{\vec q^{~2}}
\right) + \frac{1}{4}\ln\left( \frac{\vec q_2^{~2}}{\vec q^{~2}}
\right)\ln\left( \frac{\vec q_2^{~\prime 2}}{\vec q^{~2}}
\right) + \frac{1}{4}\ln^2\left( \frac{\vec q_1^{~2}}{\vec q
_2^{~2}}\right) \right] \right.
$$
\begin{equation}
\left. - \frac{\vec q_1^{~2}\vec q_2^{~\prime 2} + \vec q_2^{~2}
\vec q_1^{~\prime 2}}{2\vec k^{~2}}\ln^2\left( \frac{\vec q
_1^{~2}}{\vec q_2^{~2}}\right) + \frac{\vec q_1^{~2}\vec q
_2^{~\prime 2} - \vec q_2^{~2}\vec q_1^{~\prime 2}}{\vec k
^{~2}}\ln\left( \frac{\vec q_1^{~2}}{\vec q_2^{~2}}\right)
\left( \frac{11}{6} - \frac{1}{4}\ln\left( \frac{\vec q_1^{~2}\vec q
_2^{~2}}{\vec k^{~4}}\right) \right) \right\} + \left\{ \vec q_i \leftrightarrow \vec q_i^{~\prime} \right\}
\label{z12}
\end{equation}
while the last cannot be written through elementary functions. It has the
following integral representation: 
$$
{\cal K}_{3}^{(0)}=\left\{ \frac{1}{2}\left[ \vec q^{~2}\left( \vec k^{~2} - \vec q
_1^{~2} - \vec q_2^{~2} \right) + 2\vec q_1^{~2}\vec q_2^{~2} - \vec q
_1^{~2}\vec q_2^{~\prime 2} - \vec q_2^{~2}\vec q
_1^{~\prime 2} + \frac{\vec q_1^{~2}\vec q_2^{~\prime 2} - \vec q
_2^{~2}\vec q_1^{~\prime 2}}{\vec k^{~2}}\left( \vec q
_1^{~2} - \vec q_2^{~2} \right) \right] \right.
$$
\begin{equation}
\left.\times\int_0^1\frac{dx}{(\vec q_1(1-x) + \vec q_2x)^2}\ln
\left( \frac{\vec q_1^{~2}(1-x) + \vec q_2^{~2}x}{\vec k^{~2}x(1-x)} \right) \right\}
+ \left\{ \vec q_i \leftrightarrow \vec q_i^{~\prime} \right\}~.
\label{z13}
\end{equation}

Knowing the ``real'' kernel only in the limit $\epsilon \rightarrow 0$,
we can think about the check by the bootstrap equation of only the terms in
the gluon trajectory non-vanishing for $\epsilon \rightarrow 0$.
But we shall see that even for this purpose the knowledge of
the kernel (\ref{z10}) is not sufficient. The reason comes from the
singular behaviour of the integration measure in the bootstrap
condition (\ref{z1}).
Let us note that terms with the trajectory contribution in the
formula (\ref{z8}) for the kernel can be integrated in
Eq. (\ref{z1}) in a general form, even without the knowledge of an
explicit expression for the trajectory. The matter is that the dependence of 
$\omega ^{(2)}(t)$ on $t$ is dictated by the dimension of the bare
coupling constant, so that 
\begin{equation}
\omega ^{(2)}(t)=\bar{g}^{4}(\vec q^{~2})^{2\epsilon}N_{\epsilon}~,
\label{z14}
\end{equation}
where $N_{\epsilon }$ is some coefficient depending on $\epsilon $ (whose
expression in the limit $\epsilon \rightarrow 0$ is given by
Eq. (\ref{z7})). Using the generic integral
$$
\frac{1}{(2\pi )^{D-1}}\int \frac{d^{D-2}k}{(\vec{k}\,^{2})^{1-\delta _{1}}((%
\vec{k}-\vec{q})^{2})^{1-\delta _{2}}}=
$$
\begin{equation}
\frac{2(\vec{q}\,^{2})^{\epsilon +\delta _{1}+\delta _{2}-1}}{(4\pi
)^{2+\epsilon }}\frac{B(\epsilon +\delta _{1},\epsilon +\delta _{2})\Gamma
(1-\delta _{1}-\delta _{2}-\epsilon )}{\Gamma (1-\delta _{1})\Gamma
(1-\delta _{2})}~,
\label{z15}
\end{equation}
where $B(x,y)$ is the Euler beta-function, putting $\delta _{1}=2\epsilon $, 
$\delta _{2}=0$, from Eqs. (\ref{z1}), (\ref{z3}) and (\ref{z8}) we obtain 
$$
(\vec q^{~2})^{1-3\epsilon }\int\frac{d^{D-2}q_1}{\vec q_1^{~2}\vec q
_1^{~\prime 2}}\int\frac{d^{D-2}q_2}{\vec q_2^{~2}\vec q
_2^{~\prime 2}}\frac{{\cal K}_r^{\left( 8\right) \left( 1\right)
}\left( \vec{q}_{1},\vec{q}_{2};\vec{q}\right) }{\bar{g}^{4}\pi ^{1+\epsilon
}\Gamma (1-\epsilon )}=
$$
\begin{equation}
N_{\epsilon }B(\epsilon ,\epsilon )\left[ 1+\frac{B(\epsilon ,3\epsilon )}{%
\epsilon B(1-\epsilon ,-2\epsilon )B(\epsilon ,\epsilon )}\right]~.
\label{z16}
\end{equation}
The integration measure in Eq. (\ref{z16}) is singular at zero
momenta of scattered Reggeons. Fortunately, the kernel (\ref{z10}) turns
into zero at this points~\cite{FG}, as it can be checked by direct
inspection of Eqs. (\ref{z11}) - (\ref{z13}), so that at first sight
these points could not bring additional singularities in $\epsilon $. It
would be so if integration over only one momentum was performed. But in the
region where the momenta of two Reggeons (let us say, $\vec{q}_{1}^{\,}$ and $%
\vec{q}_{2}^{\,}$) turn into zero simultaneously, being of the same order,
the kernel does not vanish (as it can be easily observed in the example of
the Born kernel (\ref{z9})). Therefore, these regions can give additional
singularities in $\epsilon $; moreover, since integration over these regions
leads to singularities, an expansion of, let us say, $(\vec{q}%
_{1}^{\,2})^{\epsilon }$ and $(\vec{q}_{2}^{\,2})^{\epsilon }$ is not
possible anymore, so that we need to know about the kernel more than that is
given by Eqs. (\ref{z10}) - (\ref{z13}).

\section{Proof of the Bootstrap}

In the limit $\epsilon \rightarrow 0$ the coefficient $N_\epsilon$
Eq. (\ref{z14}) is defined from Eq. (\ref{z7}). Our aim is to check the non-vanishing, for $\epsilon
\rightarrow 0$, term in $N_\epsilon$. Using the explicit expression for $N_\epsilon$ we can
rewrite Eq. (\ref{z16}) as 
$$
(\vec q^{~2})^{1-3\epsilon }\int\frac{d^{D-2}q_1}{\vec q_1^{~2}\vec q
_1^{~\prime 2}}\int\frac{d^{D-2}q_2}{\vec q_2^{~2}\vec q
_2^{~\prime 2}}\frac{{\cal K}_{1}+{\cal K}_{2}+{\cal K}_{3}}{%
(\pi ^{1+\epsilon }\Gamma (1-\epsilon ))^{2}} \simeq
$$
\begin{equation}
-\frac{2}{3\epsilon ^{3}}\left[ \frac{11}{3}+\left( 2\psi ^{\prime }(1)-%
\frac{67}{9}\right) \epsilon +\left( \frac{404}{27}-11\psi ^{\prime
}(1)+\psi ^{\prime \prime }(1)\right) \epsilon ^{2}\right]~,
\label{z17}
\end{equation}
where in the R.H.S. we have kept only terms singular in $\epsilon $ (that
corresponds to non-vanishing terms in $\omega ^{(2)}$). We need to
calculate the L.H.S. of Eq. (\ref{z17}) with the same accuracy.
Let us consider where the terms singular in $\epsilon $ can come from. In
general, there are two kinds of singularities: the singularity already
existing in the kernel (namely, that in the expression (\ref{z11})
for ${\cal K}_{1}$), and the singularities appearing as the results of
integration in Eq. (\ref{z17}). The last singularities, in turn,
can come from two kinds of integration regions: one is the region
$|\vec k| \rightarrow 0$, where the kernel (again ${\cal K}_{1}$,
and only this part of the kernel) is singular, and the other consists in the
regions where the transverse momenta of the two Reggeons simultaneously tend
to zero, as it was explained at the end of the preceding section. These
last regions do not overlap each others, but overlap with the region
$|\vec k| \rightarrow 0$. It is clear that the most dangerous is the term $%
{\cal K}_{1}$ of the kernel, proportional to the Born kernel, which gives
the singularities of all kinds. Nevertheless, it is not very difficult to
calculate the contribution of this part in Eq. (\ref{z17}). The
main point here is that we know this part of the kernel sufficiently well in
the singular regions.

First of all, let us understand, in which of the regions, where the momenta
of the two Reggeons are simultaneously tending to zero, the part
${\cal K}_{1}$ can contribute. Such regions are four, in general:
\begin{eqnarray}
&1)&\ \ \ |\vec q_1| \sim |\vec q_2| \rightarrow 0;\ \ \ \vec q_1^{~\prime} \simeq \vec q
_2^{~\prime} \simeq - \vec q,\ \ \ |\vec k| \rightarrow 0~, \nonumber \\
&2)&\ \ \ |\vec q_1^{~\prime}| \sim |\vec q
_2^{~\prime}| \rightarrow 0;\ \ \ \vec q
_1 \simeq \vec q_2 \simeq \vec q,\ \ \ |\vec k
| \rightarrow 0~, \nonumber \\
&3)&\ \ \ |\vec q_1| \sim |\vec q_2^{~\prime}| \rightarrow 0;\ \ \ \vec q_1^{~\prime } \simeq -
\vec q_2 \simeq \vec k \simeq - \vec q~, \nonumber \\
&4)&\ \ \ |\vec q_1^{~\prime}| \sim |\vec q
_2| \rightarrow 0;\ \ \ \vec q_1 \simeq -
\vec q_2^{~\prime} \simeq \vec k \simeq \vec q~.
\label{z18}
\end{eqnarray}
Because of the symmetry of ${\cal K}_{1}$ with respect to $%
\vec q_i \leftrightarrow \vec q_i^{~\prime}$ it is sufficient to consider only the
first and third regions. One can easily see that in the third region $%
{\cal K}_{1} \rightarrow 0$, so that only the first region remains.
However, this is the region of small $|\vec k|$, where we know the
kernel sufficiently well, so that for the calculation we do not need more
information than that shown in Eq. (\ref{z11}).

The calculation could be done (and it was done) straightforwardly, but a
more sophisticated way of calculation leads to our aim through a much more
easy way. Let us first of all put 
\begin{equation}
d\rho =\frac{1}{\pi^{2\epsilon+2}\Gamma^2\left( 1-\epsilon \right)}
\frac{d^{D-2}q_1}{\vec q_1^{~2}\vec q_1^{~\prime 2}}\frac{
d^{D-2}q_2}{\vec q_2^{~2}\vec q_2^{~\prime 2}}~.
\label{z19}
\end{equation}
We need to calculate the terms non-vanishing with $\epsilon $ in the integral
$$
J = \int d\rho f_B\left( \frac{\vec k^2}{\vec q^{~2}} \right)
^{\epsilon} \equiv \int d\rho f_B+\int d\rho f_B\left( \left( \frac{\vec
k^2}{\vec q^{~2}} \right)^{\epsilon } - 1 \right)\theta\left( \mu^2-%
\vec k^2 \right) 
$$
\begin{equation}
+ \int d\rho f_B\left( \left( \frac{\vec k^2}{\vec q^{~2}} \right)
^{\epsilon} - 1 \right)\theta\left( \vec k^2 - \mu^2 \right)~,
\label{z20}
\end{equation}
where $\theta\left(x\right) $ is the usual step function and $\mu$ is chosen in such a way that the
ratio $\mu ^2/\vec q^{~2} \sim \epsilon^n \ll 1$, $n$ being arbitrary fixed
integer number. Then last integral is of order $\epsilon$ (since we can
expand here $\left( \vec k^2/\vec q^{~2} \right)^\epsilon$ and the regions
of singularities are excluded by the $\theta $-function) and can be
neglected, whereas the first one can be easily calculated with the help of
Eq. (\ref{z15}) and gives
\begin{equation}
\int d\rho f_B \simeq \left( \vec q^{~2} \right)
^{2\epsilon -1}\left( \frac{2}{\epsilon^2}-4\psi^\prime\left(
1 \right) \right)~.
\label{z21}
\end{equation}
In the second integral we can perform the first integration over
$\vec q_1$ keeping $\vec k$ fixed. In terms of these variables one can write 
$$
\frac{f_B}{\vec q_1^{~2}\vec q_1^{~\prime 2}\vec q_2^{~2}
\vec q_2^{~\prime 2}} = \frac{1}{\vec{k}\,^{2}}\left( \frac{1}{\left( 
\vec{q}_{1}^{\,}-\vec{q}\right) ^{2}\left( \vec{q}_{1}-\vec{k}\right) ^{2}}+%
\frac{1}{\vec{q}\,_{1}^{2}\left( \vec{q}_{1}-\vec{k}-\vec{q}\right) ^{2}}%
\right) 
$$
\begin{equation}
-\frac{\vec{q}^{~2}}{\vec{q}_{1}^{\,2}\left( \vec{q}_{1}^{\,}-\vec{q}\right)
^{2}\left( \vec{q}_{1}^{\,}-\vec{k}\right) ^{2}\left( \vec{q}_{1}^{\,}-\vec{k%
}-\vec{q}\right) ^{2}}~.
\label{z22}
\end{equation}
The first two terms can be easily integrated over $\vec{q}_{_{1}}$; as for
the last term, two non-overlapping regions do contribute to the integral at
small $|\vec k|$: $|\vec q_1| \sim |\vec k|$ and $|\vec q_1 - \vec q| \sim |\vec k|$.
They give equal contributions, so it is sufficient considering
the first of them, where we can put 
\begin{equation}
\frac{\vec q^{~2}}{\vec q_1^{~2}\vec q_1^{~\prime 2}\vec q
_2^{~2}\vec q_2^{~\prime 2}} \simeq \frac{1}{\vec{q}^{\,2}\vec{q}%
_{1}^{\,2}\left( \vec{q}_{1}^{\,}-\vec{k}\right) ^{2}}~,
\label{z23}
\end{equation}
and then expand the integration region at all $\vec q_1$, due to the
convergence of the integral. In such a way we obtain
$$
\int d\rho f_{B}\left( \left( \frac{\vec{k}\,^{2}}{\vec{q}\,^{2}}\right)
^{\epsilon }-1\right) \theta \left( \mu ^{2}-\vec{k}\,^{2}\right) \simeq
$$
\begin{equation}
\frac{2}{\vec{q}\,^{2}}\int \frac{d^{D-2}k}{\pi ^{\epsilon +1}\Gamma \left(
1-\epsilon \right) }\frac{1}{\vec{k}\,^{2}}\left( \left( \frac{\vec{k}\,^{2}%
}{\vec{q}\,^{2}}\right) ^{\epsilon }-1\right) \theta \left( \mu ^{2}-\vec{k}%
\,^{2}\right) B(\epsilon ,\epsilon )\left( \left( \vec{q}\,^{2}\right)
^{\epsilon }-\left(\vec{k}\,^{2} \right)^{\epsilon }\right)~.
\label{z24}
\end{equation}
Using the relation 
\begin{equation}
\frac{d^{D-2}k}{\pi ^{\epsilon +1}\Gamma \left( 1-\epsilon \right) } =
\frac{\left( \vec k^{~2} \right)^{\epsilon}d\vec k^{~2}}{\Gamma (1-\epsilon)\Gamma (1+\epsilon)}
\label{z25}
\end{equation}
we obtain
\begin{equation}
\int d\rho f_{B}\left( \left( \frac{\vec k^{~2}}{\vec q^{~2}}\right) ^{\epsilon
}-1\right) \theta \left( \mu ^{2}-\vec k^{~2}\right) \simeq \left( \vec{q}\,^{2}\right)
^{2\epsilon -1}\left[ -\frac{4}{3\epsilon ^{2}}+\frac{8}{3}\psi ^{\prime
}\left( 1\right) \right]~.
\label{z26}
\end{equation}
This result, together with Eq. (\ref{z21}) gives
\begin{equation}
J \simeq \left( \vec q^{~2} \right)^{2\epsilon -1}\left[ \frac{2}{%
3\epsilon^2}-\frac{4}{3}\psi^\prime\left( 1 \right) \right]~,
\label{z27}
\end{equation}
and, consequently, 
$$
\left( \vec q^{~2} \right)^{1-3\epsilon}\int\frac{d^{D-2}q_1}{\vec q
_1^{~2}\vec q_1^{~\prime 2}}\int\frac{d^{D-2}q_2}{\vec q
_2^{~2}\vec q_2^{~\prime 2}}\frac{{\cal K}_{1}}{(\pi
^{1+\epsilon }\Gamma (1-\epsilon ))^{2}} \simeq
$$
\begin{equation}
-\frac{2}{3\epsilon ^{3}}\left[ \frac{11}{3}+\left( 2\psi ^{\prime }(1)-%
\frac{67}{9}\right) \epsilon +\left( \frac{404}{27}-11\psi ^{\prime
}(1)+7\psi ^{\prime \prime }(1)\right) \epsilon ^{2}\right]~.
\label{z28}
\end{equation}
This result coincides with that of the straightforward calculation which was
also performed. Comparing with Eq. (\ref{z17}) we see that the contribution of the
term proportional to the Born kernel almost ``saturates'' \ the bootstrap
condition; the only difference is that, instead of the term  $7\psi ^{\prime
\prime }(1)$ of Eq. (\ref{z28}), in the R.H.S. of Eq. (\ref{z17}) there is $\psi ^{\prime \prime }(1)$.

We now turn to the remaining contributions to the kernel, ${\cal K}_2$ and ${\cal K}_3$.
They have neither explicit singularities in $\epsilon $,
nor singular behaviour for $|\vec k| \rightarrow 0$. Therefore,
terms of order 1/$\epsilon $ can be obtained only from the regions 1) - 4)
(see Eq. (\ref{z18})), where the momenta of the two Reggeons tend
to zero being of the same order. It is easy to see from Eqs. (\ref{z12}) and (\ref{z13}) that
in the regions 3) and 4) both ${\cal K}_2^{(0)}$ and ${\cal K}_3^{(0)}$ turn into
zero. Using the symmetry of the kernel with respect to $\vec q_i \leftrightarrow \vec q_i^{~\prime}$,
we can consider only the first region. Here we have
$$
{\cal K}_2^{(0)} \simeq \vec q^{~2}\left[ \frac{11}{6}\left( \ln\left( \frac{\vec q_1^{~2}\vec q_2^{~2}}
{\vec k^{~4}} \right) + \frac{\vec q_1^{~2} - \vec q_2^{~2}}{\vec k^{~2}}\ln\left( \frac{\vec q_1^{~2}}
{\vec q_2^{~2}} \right) \right) + \left( \frac{1}{4} - \frac{\vec q_1^{~2} + \vec q_2^{~2}}{2\vec k^{~2}}
\right)\ln^2\left( \frac{\vec q_1^{~2}}{\vec q_2^{~2}} \right) \right.
$$
\begin{equation}\label{z29}
\left. - \frac{1}{4}\frac{\vec q_1^{~2} - \vec q_2^{~2}}{\vec k^{~2}}\ln\left( \frac{\vec q_1^{~2}}
{\vec q_2^{~2}} \right)\ln\left( \frac{\vec q_1^{~2}\vec q_2^{~2}}{\vec k^{~4}} \right) \right]~,
\end{equation}
\begin{equation}\label{z30}
{\cal K}_3^{(0)} \simeq \frac{\vec q^{~2}}{2}\left[ \vec k^{~2} - \vec q_1^{~2} - \vec q_2^{~2} +
\frac{2\left( \vec q_1^{~2}(\vec q_2\vec k) - \vec q_2^{~2}(\vec q_1\vec k) \right)}{\vec k^{~2}} \right]I~,
\end{equation}
where
\begin{equation}\label{z31}
I = \int_0^1\frac{dx}{(\vec q_1(1-x) + \vec q_2x)^2}\ln\left( \frac{\vec q_1^{~2}(1-x) + \vec q_2^{~2}x}
{\vec k^{~2}x(1-x)} \right)~.
\end{equation}
If we fix $\vec k$ and perform the integration over $\vec q_1$, then it is convergent at $|\vec q_1| \sim |\vec k|$
since
\begin{equation}\label{z32}
d\rho \simeq \frac{1}{\pi^{2\epsilon+2}\Gamma^2\left( 1-\epsilon \right)}
\frac{d^{D-2}k}{\vec q^{~4}}\frac{d^{D-2}q_1}{\vec q_1^{~2}(\vec q_1 - \vec k)^2}~.
\end{equation}
Therefore with such order of integration at fixed small $\vec k$ we can integrate over all $\vec q_1$ and
nevertheless use the expressions (\ref{z29}) - (\ref{z31}).

All integrals of the terms entering into Eq. (\ref{z29}) for ${\cal K}_2^{(0)}$ can be taken using Eq. (\ref{z15})
and its derivatives with respect to $\delta_1$ and $\delta_2$. All
the integrals of this part can be calculated for arbitrary $\epsilon$. We are interested in the limit
$\epsilon \rightarrow 0$, but in the region of arbitrary small $\vec k$, so that we cannot expand
$(\vec k^{~2})^\epsilon$ in powers of $\epsilon$. In this limit we obtain
\begin{equation}\label{z33}
\int\frac{d^{D-2}q_1}{\vec q_1^{~2}(\vec q_1 - \vec k)^2}\frac{{\cal K}_2^{(0)}}{\pi^{1+\epsilon}\Gamma
(1-\epsilon )} \simeq \frac{\vec q^{~2}}{\vec k^{~2}}\left( \vec k^{~2} \right)^\epsilon
3\psi^{\prime\prime}(1)~.
\end{equation}
The integration of Eq. (\ref{z30}) for ${\cal K}_3^{(0)}$ is not straightforward. We use the representation
$$
I = \int_0^1dx\int_1^\infty\frac{dz}{z}\frac{1}{\left( \vec q_1 - x\vec k \right)^2 + zx(1-x)\vec k^{~2}} =
$$
\begin{equation}\label{z34}
\int_0^1dx\int_1^\infty\frac{dz}{z}\frac{1}{x(\vec q_1 - \vec k)^2 + (1-x)\vec q_1^{~2} + (z-1)x(1-x)\vec k^{~2}}~.
\end{equation}
After this the integration over $\vec q_1$ is reduced to integrations of terms with two denominators, which are
performed using the standard Feynman parametrization. Remaining integrations over $x$, $z$ and the Feynman
parameter are rather tedious, although not very complicated. The result for $\epsilon \rightarrow 0$ is the same
as for ${\cal K}_2^{(0)}$:
\begin{equation}\label{z35}
\int\frac{d^{D-2}q_1}{\vec q_1^{~2}(\vec q_1 - \vec k)^2}\frac{{\cal K}_3^{(0)}}{\pi^{1+\epsilon}\Gamma
(1-\epsilon )} \simeq \frac{\vec q^{~2}}{\vec k^{~2}}\left( \vec k^{~2} \right)^\epsilon
3\psi^{\prime\prime}(1)~.
\end{equation}
Again we do not expand $\left( \vec k^{~2} \right)^\epsilon$, because at the subsequent integration over $\vec k$
the region where $\epsilon\left| \ln\vec k^{~2}  \right| \sim 1$ does contribute. Note that appearance of the factors
$\left( \vec k^{~2} \right)^\epsilon$ in Eqs. (\ref{z33}) and (\ref{z35}) is evident without calculations: they are
dimensional factors and in the integrals (\ref{z33}) and (\ref{z35}) we have only one dimensional parameter.

We see that scale factors can be important even in calculation of terms singular in $\epsilon$ which come from the
nonsingular contributions ${\cal K}_2$ and ${\cal K}_3$ in Eq. (\ref{z17}). But we know ${\cal K}_2$ and ${\cal K}_3$
only in the limit $\epsilon \rightarrow 0$ at fixed Reggeon momenta, where scale factors are put equal to 1
(see Eqs. (\ref{z12}) and (\ref{z13})). The problem therefore is to restore the scale factors. Fortunately, we need
to know
the scale factors only in the first of regions (\ref{z18}), where we have only two essentially different scales since
we have $|\vec q_1| \sim |\vec q_2| \sim |\vec k|$ from one side and
$|\vec q_1^{~\prime}| \simeq |\vec q_2^{~\prime}| \simeq |\vec q|$ from the other. Even without calculations
it is clear that the relevant scale should be $|\vec k|$, since the scale appears as a result of integration over
transverse momenta (transverse momenta of gluons produced in the Reggeon-Reggeon collision for the two-gluon
contribution to the kernel, transverse momenta of virtual gluons in the radiative correction to the
Reggeon-Reggeon-gluon (RRG) vertex for the one-gluon contribution), and for
$|\vec q_1| \sim |\vec q_2| \sim |\vec k|$ in the
essential region of the integration these momenta should be of the same order. This conclusion is confirmed by
direct inspection of the kernel. Fortunately, for the two-gluon contribution this inspection can be done
straightforwardly, since there is the explicit expression for this contribution for arbitrary $D$~\cite{FG}. The
one-gluon contribution consists of two pieces: in one of them radiative corrections are contained in the RRG
vertex with momenta $q_1$ and $q_2$, in the other with momenta $q_1^\prime$ and $q_2^\prime$. It follows
from the kinematical structure of the RRG vertex that in the region 1) only the first piece does contribute to
${\cal K}_2$ (${\cal K}_3$ comes totally from the two-gluon production). The evident scale for this piece is
$|\vec k|$. Therefore, in the first of regions (\ref{z18}) we have
\begin{equation}\label{z36}
{\cal K}_2 = \left( \vec k^{~2} \right)^\epsilon{\cal K}_2^{(0)}~,\ \ \ {\cal K}_3= \left( \vec k^{~2} \right)
^\epsilon{\cal K}_3^{(0)}~.
\end{equation}
Using Eqs. (\ref{z33}) and (\ref{z35}), and taking into account the region 2) by doubling the results we obtain
$$
\vec q^{~2}\int\frac{d^{D-2}q_1}{\vec q_1^{~2}\vec q_1^{~\prime 2}}\int\frac{d^{D-2}q_2}{\vec q
_2^{~2}\vec q_2^{~\prime 2}}\frac{{\cal K}_2 + {\cal K}_3}{(\pi^{1+\epsilon}\Gamma(1-\epsilon))^2} \simeq
$$
\begin{equation}\label{z37}
12\psi^{\prime\prime}(1)\int\frac{d^{D-2}k}{\pi^{1+\epsilon}\Gamma(1-\epsilon)}\left( \vec k^{~2}
\right)^{2\epsilon-1}\theta\left( \vec q^{~2} - \vec k^{~2} \right) \simeq \frac{4\psi^{\prime\prime}(1)}
{\epsilon}~.
\end{equation}
Putting Eqs. (\ref{z28}) and (\ref{z37}) into Eq. (\ref{z17}) we see that the bootstrap condition is satisfied.

\section{Conclusions}

We have proved the fulfillment of the first bootstrap condition in the form of Eq. (\ref{z1}) in the limit of the space-time
dimension $D$ tending to its physical value $D=4$. All terms in the two-loop contribution
to the gluon trajectory and in the NLO correction to the octet BFKL kernel, non-vanishing in this limit, are involved in
this bootstrap. Therefore,
the performed verification of the bootstrap gives us not only a powerful confirmation of the gluon Reggeization,
but at the same time a stringent test of the calculations of the trajectory and the kernel.

Now we have practically no doubts on the gluon Reggeization in the NLO, as well as on the calculations of gluon
trajectory and kernel. Nevertheless it would be interesting to verify if the first bootstrap condition is satisfied
for arbitrary space-time dimension $D$. That is known to be true for the quark part of the kernel and for the
second bootstrap
condition in the cases of quark and gluon impact factors. It will become possible to do this verification after the
calculation of the
one-loop correction to the Reggeon-Reggeon-gluon vertex for arbitrary $D$, which is in progress now~\cite{FFP00}.
Another interesting possibility is to check the so-called ``first strong bootstrap condition'' for the kernel suggested
by Braun and Vacca~\cite{StBoot}. This condition is derived from the requirement that the particle-Reggeon 
scattering amplitudes have a Reggeized form and it is satisfied for the quark part of the kernel, as well as the
analogous
condition for impact factors in the quark and gluon cases, although the role of the strong bootstrap conditions in
the BFKL approach is not completely understood.

\underline{Acknowledgment:} Two of us (V.S.F. and M.I.K.) thank the Dipartimento di Fisica della Universit\`a della
Calabria for the warm hospitality while a part of this work was done. V.S.F. acknowledges also the financial support of
the Istituto Nazionale di Fisica Nucleare.

\end{document}